\documentclass[prb,twocolumn,amsmath,amssymb,nofootinbib,floatfix,superscriptaddress]{revtex4}

\usepackage{graphicx,bm}
\usepackage[normalem]{ulem} 

\makeatletter
\def\graphicscale{\twocolumn@sw{0.3}{0.4}}
\def\graphicthreescale{\twocolumn@sw{0.3}{0.4}}

\begin{document}

\title{Multicritical point of the three-dimensional
  ${\mathbb Z}_2$
  gauge Higgs model}

\author{Claudio Bonati} \affiliation{Dipartimento di Fisica
  dell'Universit\`a di Pisa and INFN Largo Pontecorvo 3, I-56127 Pisa,
  Italy}

\author{Andrea Pelissetto}
\affiliation{Dipartimento di Fisica dell'Universit\`a di Roma Sapienza
        and INFN Sezione di Roma I, I-00185 Roma, Italy}

\author{Ettore Vicari} 
\affiliation{Dipartimento di Fisica dell'Universit\`a di Pisa
        and INFN Largo Pontecorvo 3, I-56127 Pisa, Italy}

\date{\today}

\begin{abstract}
  We investigate the multicritical behavior of the three-dimensional
  ${\mathbb Z}_2$ gauge Higgs model, at the multicritical point (MCP)
  of its phase diagram, where one first-order transition line and two
  continuous Ising-like transition lines meet. The duality properties
  of the model determine some features of the multicritical behavior
  at the MCP located along the self-dual line. Moreover, we argue that
  the system develops a multicritical $XY$ behavior at the MCP, which
  is controlled by the stable $XY$ fixed point of the
  three-dimensional multicritical Landau-Ginzburg-Wilson field theory
  with two competing scalar fields associated with the continuous
  ${\mathbb Z}_2$ transition lines meeting at the MCP.  This implies
  an effective enlargement of the symmetry of the multicritical modes
  at the MCP, to the continuous group O(2).  We also provide some
  numerical results to support the multicritical $XY$ scenario.
\end{abstract}

\maketitle

\section{Introduction}
\label{intro}

\begin{figure}[tbp]
\includegraphics[width=0.95\columnwidth, clip]{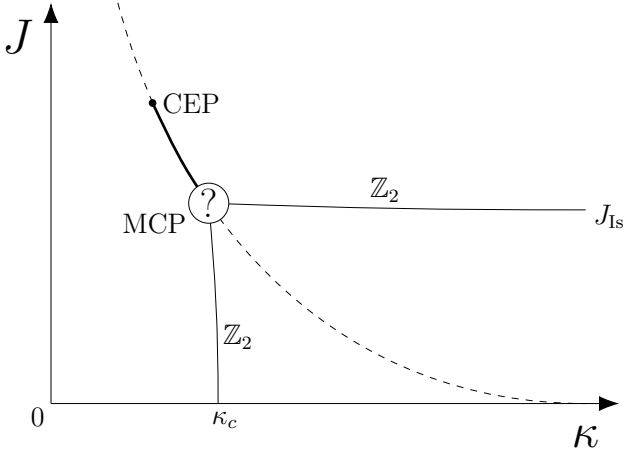}
\caption{Sketch of the phase diagram of the 3D ${\mathbb Z}_2$ gauge
  Higgs model (\ref{HiggsH}). The dashed line is the self-dual line,
  cf. Eq.~(\ref{selfdual}), the thick line corresponds to first-order
  transitions on the self-dual line, extending for a finite interval.
  The two lines labelled ``${\mathbb Z}_2$" are related by duality,
  cf. Eq.~(\ref{dualitymap}), and correspond to Ising-like continuous
  transitions. They end at $J = J_{\rm Is} \approx 0.22165$,
  $\kappa=\infty$ and at $J =0$, $\kappa = \kappa_c \approx 0.76141$.
  The three lines are conjectured to meet at a multicritical point
  (MCP) on the self-dual line, at $[\kappa^\star\approx
    0.7525,J^\star\approx 0.2258]$. We argue in the paper that the
  multicritical behavior belongs to the $XY$ universality class.  The
  other endpoint of the first-order transition line should give rise
  to a critical endpoint (CEP).  }
\label{phadia}
\end{figure}

The three-dimensional (3D) ${\mathbb Z}_2$ gauge Higgs model is one of
the simplest gauge theories with matter fields, that shows a
nontrivial phase diagram characterized by the presence of a
topological phase, see, e.g.,
Refs.~\onlinecite{Wegner-71,BDI-74,BDI-75,FS-79,Kogut-79,JSJ-80,HL-91,
  GGRT-03,Kitaev-03,VDS-09,TKPS-10,GHMS-11,DKOSV-11,WDP-12,
  Fradkin-book,Sachdev-19,SSN-21,Grady-21,HSAFG-21}.  The model can
also be related to the quantum two-dimensional toric model in the
presence of external {\em magnetic} fields, by a quantum-to-classical
mapping~\cite{Wegner-71,Kitaev-03,TKPS-10}, and to a statistical
ensemble of membranes~\cite{HL-91,GHMS-11}.

A notable feature of the model~\cite{Wegner-71,FS-79,Kogut-79} is the
existence of a duality transformation, which relates the free energy
at different points of the phase
diagram~\cite{Wegner-71,BDI-75,FS-79,Kogut-79}. A particular line in
the phase diagram, which will play an important role in the following,
is the self-dual line which is left invariant by the duality
transformation.  In Fig.~\ref{phadia} we sketch the phase diagram of
the model, in the space of the Hamiltonian parameters [they are
  defined in Eq.~(\ref{HiggsH})].  It presents a topologically ordered
deconfined phase, delimited by two continuous Ising transition lines
that are related by duality.  In the context of two-dimensional
quantum systems, such a topological ordered phase is realized in
${\mathbb Z}_2$ spin
liquids~\cite{RC-89,Kivelson-89,RS-91,Wen-91,SF-00,MSF-01}, which is
the phase of matter realized by the toric code~\cite{Kitaev-03}.
Moreover, the 3D ${\mathbb Z}_2$ gauge Higgs model presents a
first-order transition line running along the self-dual line, for a
limited range of the Hamiltonian
parameters~\cite{TKPS-10,GGRT-03,JSJ-80}.

The available numerical results are consistent with the existence of a
multicritical point (MCP), where the first-order transition line and
the two continuous Ising transition lines meet, see, e.g.,
Refs.~\onlinecite{TKPS-10,SSN-21}.  Assuming the existence of the MCP, an
interesting question concerns the nature of the multicritical
behavior. This issue has been recently investigated in
Ref.~\onlinecite{SSN-21}, which reported apparently puzzling results,
leading to estimates of the critical exponents that are substantially
consistent with those of the $XY$ universality class. This may suggest
that the multicritical behavior at the MCP is controlled by the 3D
$XY$ fixed point, with an effective enlargement of the symmetry of the
multicritical modes to the continuous O(2) group. This scenario was
considered unlikely in Ref.~\onlinecite{SSN-21}, because of the unclear
relationship between the multicritical $XY$ behavior and the mutual
statistics of the condensing
quasiparticles~\cite{TKPS-10,VDS-09,GM-12,Burnell-18} along the two
distinct Ising transition lines meeting at the MCP.  These mutual
statistics do not affect critical exponents on the Ising lines,
because only one of the two excitations is massless on them,
but both excitations must become massless at the MCP.  Therefore, it is
not clear how their competition can give rise to the effective
enlargement of the symmetry at the MCP, as required by the $XY$
universality class.

In this paper we investigate the multicritical behavior at the MCP.
We argue that the multicritical behavior is controlled by the stable
$XY$ fixed point of the 3D multicritical Landau-Ginzburg-Wilson (LGW)
field theory with two competing scalar fields associated with the
${\mathbb Z}_2$ transition lines meeting at the
MCP~\cite{LF-72,FN-74,NKF-74,PV-02,CPV-03}. Duality properties play a
crucial role for the realization of the multicritical $XY$ scenario,
which implies an effective enlargement of the symmetry of the
multicritical modes, to the continuous symmetry group O(2).  To
provide further support to this scenario, we also report some
numerical finite-size scaling (FSS) analyses of data from Monte Carlo
(MC) simulations.

The paper is organized as follows.  In Sec.~\ref{model} we present the
3D lattice ${\mathbb Z}_2$ gauge Higgs model, and summarize the known
features of its phase diagram. In Sec.~\ref{multicr} we discuss the
multicritical theory appropriate for the MCP, and apply the
multicritical LGW field theory to predict a multicritical $XY$
behavior. In Sec.~\ref{numres} we report some numerical results
supporting the multicritical $XY$ scenario, obtained by FSS analyses
of MC simulations. Finally in Sec.~\ref{conclu} we draw our
conclusions.

\section{The ${\mathbb Z}_2$ gauge Higgs model}
\label{model}

\subsection{Hamiltonian and duality transformations}
\label{modelH}

We consider a lattice gauge model with ${\mathbb Z}_2$ gauge
invariance defined on a cubic 3D lattice with periodic boundary
conditions.  The fundamental variables are Ising spins $s_{\bm x}=\pm
1$ defined on the lattice sites and Ising spins $\sigma_{{\bm
    x},\mu}=\pm 1$ defined on the bonds ($\sigma_{{\bm x},\mu}$ is
associated with the bond starting from site ${\bm x}$ in the $\mu$
direction, $\mu=1,2,3$). The model is defined by the lattice
Hamiltonian~\cite{Wegner-71,FS-79,Kogut-79}
\begin{eqnarray}
  &&  H = - J \sum_{{\bm x},\mu}
  s_{\bm x} \, \sigma_{{\bm x},\mu} \,
  s_{{\bm x}+\hat{\mu}}
  - \kappa \sum_{{\bm x},\mu>\nu} \Pi_{{\bm
      x},\mu\nu}\,,
\label{HiggsH}\\
&&\Pi_{{\bm x},\mu\nu}=
 \sigma_{{\bm
      x},\mu} \,\sigma_{{\bm x}+\hat{\mu},\nu} \,\sigma_{{\bm
      x}+\hat{\nu},\mu} \,\sigma_{{\bm x},\nu}\,.
\label{plaquette}
\end{eqnarray}
The corresponding partition function and free-energy density are
\begin{equation}
  Z = \sum_{\{s,\sigma\}} e^{-\beta H(J,\kappa)}\,, \qquad
  F(J,\kappa) = - {T\over L^d} \ln Z\,,
  \label{partfuncmodel}
\end{equation}
where $\beta=1/T$ is the inverse temperature, and $L^d$ is the volume
of the system. This paper only consider three-dimensional systems, and
therefore $d=3$. However, when arguments are independent of the space
dimension, we keep $d$ generic.  In the following, energies are
measured in units of $T$, which is equivalent to fix $\beta=1$ in
Eq.~\eqref{partfuncmodel}.

The model can be simplified by considering the so-called unitary
gauge. Indeed, the site variables $s_{\bm x}$ can be eliminated by
redefining $\sigma_{{\bm x},\mu}$ as
\begin{equation}
  s_{\bm x} \, \sigma_{{\bm x},\mu} \, s_{{\bm x}+\hat{\mu}}\,\to\,
  \sigma_{{\bm x},\mu} \,.
  \label{gaugeun}
\end{equation}
Correspondingly, the partition function can be written as
\begin{eqnarray}
&& Z = \sum_{\{\sigma\}} e^{-H_{\rm ug}(J,\kappa)}\,,\label{zhiggs}\\ 
&& H_{\rm ug} = - J \sum_{{\bm x},\mu} \sigma_{{\bm x},\mu} - \kappa
  \sum_{{\bm x},\mu>\nu} \Pi_{{\bm x},\mu\nu}\,.
\label{HiggsHug}
\end{eqnarray}

An important property of the 3D lattice ${\mathbb Z}_2$ gauge Higgs
model is the existence of a duality mapping~\cite{BDI-75} between the
Hamiltonian parameters, that leaves the partition function unchanged,
modulo a regular function of the parameters. If
\begin{eqnarray}
  \left(J^\prime, \kappa^\prime\right)=
  \left( -{1\over 2} {\rm ln}\,{\rm
      tanh}\,\kappa\,, -{1\over 2} {\rm ln}\,{\rm tanh}\, J \right)\,,
  \label{dualitymap}
\end{eqnarray}
we have \cite{BDI-75}
\begin{equation}
  F(J^\prime, \kappa^\prime) = F(J, \kappa) - {3\over 2}
  \ln[\sinh(2J)\sinh(2\kappa)]\,.
\label{dualityZ}
\end{equation}
One can also define a self-dual line,
\begin{equation}
D(J,\kappa) = J - J^\prime = 
   J + {1\over 2} {\rm ln}\,{\rm tanh}\,\kappa = 0\,,
\label{selfdual}
\end{equation}
where the duality transformation maps the model into itself, i.e.
$J^\prime = J$ and $\kappa^\prime = \kappa$.  Note that $D(J,\kappa)$
is odd under the duality mapping $(J,\kappa) \to
(J^\prime,\kappa^\prime)$, i.e., $D(J,\kappa) = -
D(J^\prime,\kappa^\prime)$.

\subsection{The phase diagram}
\label{phasdiagr}

Some features of the phase diagram are well established, see, e.g.,
Refs.~\onlinecite{FS-79,TKPS-10,SSN-21}.  A sketch of the phase diagram is
shown in Fig.~\ref{phadia}.  For $\kappa\to\infty$ an Ising transition
occurs at~\cite{FXL-18} $J_{\rm Is} = 0.221654626(5)$.  By duality, in
the pure ${\mathbb Z}_2$ gauge model a transition occurs in the
corresponding point, $J=0$ and
\begin{equation}
\kappa_c = -{1\over 2} {\rm ln}\,{\rm tanh}\,J_{\rm Is} = 
  0.761413292(11)\,.
  \label{z2gaugecr}
\end{equation}
Two Ising-like continuous transition lines, related by the duality
transformation (\ref{dualitymap}), start from these
points~\cite{FS-79} and intersect along the self-dual
line~\cite{SSN-21}.  Moreover, some numerical
studies~\cite{TKPS-10,GGRT-03} have provided evidence of first-order
transitions along the self-dual line, in the relatively small interval
\begin{equation}
0.688 \lesssim \kappa \lesssim 0.753\,,\qquad
0.258\gtrsim J \gtrsim 0.226\,.
  \label{foint}
  \end{equation}
Since the first-order transition line is limited to an interval along
the self-dual line, there are only two phases, separated by the two
continuous transition lines, see Fig.~\ref{phadia}.  For small $J$ and
large $\kappa$ there is a topological deconfined phase. The remaining
part of the phase diagram corresponds to a single phase that extends
from the disordered small-$J,\kappa$ region to the whole large-$J$
region.  In particular, no phase transition occurs along the line
$\kappa=0$, where the model (\ref{HiggsHug}) becomes trivial.

A natural conjecture is that the first-order and the two continuous
Ising transition lines meet at the same point located along the
self-dual line, giving rise to a multicritical point (MCP).  Numerical
results~\cite{SSN-21,TKPS-10} are consistent with this conjecture.  In
particular, Ref.~\onlinecite{SSN-21} reported evidence of a critical
transition point along the self-dual line---we identify it with the
MCP---with critical parameters $\kappa^\star \approx 0.7526$ and
$J^\star\approx 0.2257$.  The corresponding critical exponents are
close to, and substantially consistent with, those associated with the
$XY$ universality class~\cite{PV-02,CPV-03,CHPV-06,HV-11}.  In spite
of these results, Ref.~\onlinecite{SSN-21} considered an $XY$ multicritical
behavior unlikely.  In the paper, we rediscuss the issue, and give
additional theoretical and numerical arguments that support the
hypothesis that the MCP belongs to the $XY$ universality class.

We finally note that the first-order transition line starting from the
MCP ends at $J\approx 0.258$ and $\kappa\approx 0.688$.  We expect
this endpoint to correspond to a continuous transition, likely
belonging to the Ising universality class.

\section{Multicritical behavior}
\label{multicr}

As discussed above, the phase diagram of the lattice ${\mathbb Z}_2$
gauge Higgs model shows a MCP, where a first-order and two continuous
transition lines meet (this MCP is usually called
bicritical~\cite{LF-72,FN-74,NKF-74}).  In the following, we first
discuss the expected behavior of the model close to the MCP, on the
basis of the renormalization-group (RG) theory. Then, we discuss a
a LGW field theory characterized by two interacting local real scalar
fields~\cite{LF-72,FN-74,NKF-74,CPV-03}, which may describe the 
multicritical behavior.

\begin{figure}[tbp]
\includegraphics[width=0.95\columnwidth, clip]{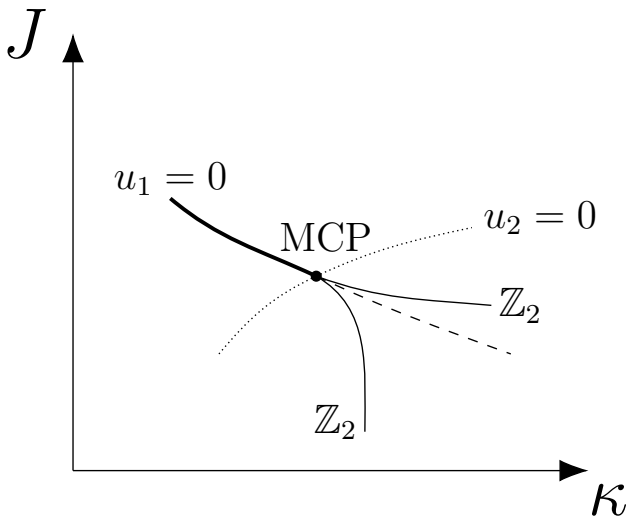}
\caption{Sketch of the phase diagram close to the MCP. We report the
  first-order transition line (thick line), the self-dual line (dashed
  line), the two continuous transition lines (continuous lines), and
  the line (dotted line) where $u_2 = 0$. The line $u_1 = 0$ coincides
  with the self-dual line.  }
\label{phadiabis}
\end{figure}

\subsection{Multicritical scaling theory}
\label{sec3A}

At a MCP, the singular part of the free-energy density
can be written as
\begin{equation}
F_{\rm sing}(J,\kappa,L) = L^{-d} {\cal F}(\{u_i L^{y_i}\})\,,
\label{freeen}
\end{equation} 
where $u_i$ are the nonlinear scaling fields and the RG exponents
$y_i$ are ordered so that
\begin{equation}
y_1>y_2>y_3>y_4 > ...\,,
\label{defyi}
\end{equation}
In the present model, we expect two relevant RG
perturbations. Therefore, $y_1$ and $y_2$ are positive, and the
corresponding scaling fields $u_1$ and $u_2$ vanish at the MCP. The
exponents $y_i$ with $i\ge 3$ are instead negative and control the
corrections to the multicritical behavior. All the scaling fields
$u_i$ are analytic functions of the model parameters $J$ and
$\kappa$. In the infinite-volume limit and neglecting subleading
corrections, we can rewrite the singular part of the free energy
density as
\begin{eqnarray}
  &&F_{\rm sing}(J,\kappa) = |u_2|^{d/y_2} {\cal F}_\pm (X)\,,
  \label{freeen2}\\
  && X \equiv
  u_1 |u_2|^{-\phi}\,, \qquad \phi\equiv {y_1/y_2}>1\,,
\label{phidef}
\end{eqnarray} 
where the functions ${\cal F}_\pm(X)$ apply to the parameter regions
in which $\pm u_2 > 0$, and $\phi$ is the so-called crossover exponent
associated with the MCP.  Close to the MCP, the transition lines
follow the equation
\begin{equation}
X = u_1 |u_2|^{-\phi} = {\rm const}\,,
\label{traju1u2}
\end{equation}
with a different constant for each transition line.  Since $\phi > 1$,
they are tangent to the line $u_1 = 0$.

The duality mapping (\ref{dualitymap}), and in particular
Eq.~(\ref{dualityZ}), implies the relation
\begin{equation}
  F_{\rm sing}(J^\prime,\kappa^\prime) = F_{\rm sing}(J,\kappa)\,.
  \label{fsingdua}
  \end{equation}
Then, if we set
\begin{equation}
u^\prime_1 =
u_1(J^\prime,\kappa^\prime)\,,\quad u^\prime_2 =
u_2(J^\prime,\kappa^\prime)\,,
\label{u1u2p}
\end{equation}
using Eq.~(\ref{freeen2}) we obtain the equality
\begin{equation}
   |u_2|^{d/y_2}{\cal F}_\pm (u_1 |u_2|^{-\phi}) = 
   |u_2^\prime|^{d/y_2}{\cal F}_\pm (u_1^\prime |u_2^\prime|^{-\phi})\,.
\label{eqSF-freeenergy}
\end{equation}
Since along the self-dual line $u_1=u_1^\prime=0$, this relation can only be
satisfied if $|u_2| = |u_2^\prime|$. If we then expand the scaling
function ${\cal F}_\pm(X)$ in powers of $X$, Eq.~(\ref{eqSF-freeenergy})
implies $u_1^m = (u_1^\prime)^m$ for all values of $m$ such that the
derivative
\begin{equation}
   {\cal F}_{m} = \left. {\partial^{m} {\cal F}(X)\over \partial X^m}
      \right \vert_{X=0}
  \label{deffm} 
\end{equation}
is nonvanishing. This condition can be satisfied only if $u_1$ changes
at most by a sign under duality. As we shall argue below, $u_1$ is odd
under duality, i.e., $u_1^\prime = -u_1$. In this case, we should
additionally require ${\cal F}_m = 0$ for any odd $m$: the functions
${\cal F}_\pm(X)$ are even in $X$.
    
To show that $u_1$ is odd, note that, as discussed in
Sec.~\ref{phasdiagr}, the first-order transition line runs along the
self-dual line (\ref{selfdual}) ending at the MCP, located at
\begin{equation}
  J = J^\star,\quad \kappa = \kappa^\star =
    -{1\over 2} {\rm ln}\,{\rm tanh}\,J^\star\,.
  \label{tcpco}
  \end{equation}
This transition line is expected to coincide~\cite{LF-72,FN-74,NKF-74}
with the line $u_1 = 0$, close to the MCP. Since the self-dual line is
given by $D(J,\kappa) = 0$, we can make the identification
\begin{eqnarray}
u_1 = D(J,\kappa)\,,
\label{nonlinu1}
\end{eqnarray}
close to the MCP. As noted before, $D(J,\kappa)$ is odd under duality.
The scaling field $u_2$ is then necessarily even under duality and is
therefore given by
\begin{equation}
   u_2(J,k) = - J + J^\star + 
{1\over 2} \ln {\tanh \kappa^{\phantom{\star}} \over \tanh\kappa^\star}\,.
\label{u2def}
\end{equation}
The scaling fields can be straightforwardly linearized obtaining
\begin{eqnarray}
&&u_1 \approx   \Delta J + c\, \Delta \kappa\,, \qquad
u_2 \approx - \Delta J + c \,\Delta \kappa\,,
\label{ulinear}
\end{eqnarray}
where
\begin{eqnarray}
&&\Delta J = J-J^\star\,,\qquad\Delta \kappa = \kappa-\kappa^\star\,,
\nonumber\\
&&c = \sinh(2J^\star) = {1\over \sinh(2\kappa^\star)} \approx 0.467\,.
\label{defcconstant}
\end{eqnarray}
In terms of $u_1$ and $u_2$, close to the MCP the first-order
transition line corresponds to $ X = 0$, $u_2<0$.  The two continuous
transition lines are defined by $X = \pm k$ with $u_2 > 0$.

Using the above results we can also predict how the latent heat
$\Delta_h$ vanishes along the first-order transition line when
approaching the MCP.  A straightforward scaling
argument~\cite{CPPV-04} gives
\begin{equation}
  \Delta_t \sim |u_2|^\theta\,,\qquad \theta = {d-y_1\over y_2}\,.
  \label{latheat}
  \end{equation}
Note that this scaling behavior is the same as that of the
magnetization $M$ at the Ising transition, with the correspondence
$y_1=y_h$ and $y_2=1/\nu$: $M$ indeed vanishes as $M\sim
|T-T_c|^\beta$ with $\beta = (d-y_h)\nu$, see, e.g.,
Ref.~\onlinecite{PV-02}.

\subsection{Scaling of the energy cumulants }
\label{sec3B}

Due to the fact that we are considering a lattice gauge theory, and
therefore that we cannot easily access the order parameters associated
with the phase transitions, we focus on the multicritical behavior of
the energy operators.  We define
\begin{eqnarray}
H_J &=& \sum_{x\,\mu} \sigma_{x,\mu}\,, \qquad
H_\kappa = \sum_{x\,\mu>\nu} \Pi_{x,\mu\nu}\,,\label{hamjk}\\
H &=& -J H_J - \kappa H_\kappa\,.\nonumber
\end{eqnarray} 
We consider the cumulants 
\begin{equation}
     C_{nm} = - L^d {\partial^{n+m} \over \partial J^n \partial
       \kappa^m} F(J,\kappa,L)\,,
\label{cumulant-def}
\end{equation}
where $F$ is the free-energy density.  For $n+m\le 3$, $C_{nm} =
M_{nm}$, where $M_{nm}$ are the central moments defined by
\begin{equation}
  M_{nm} = \langle (H_J - E_J)^n (H_\kappa - E_\kappa)^m \rangle\,, 
\label{nmomdef}
\end{equation}
with $E_J = \langle H_J \rangle$ and $E_\kappa = \langle H_\kappa
\rangle$.  For $n+m\ge 4$, central moments and cumulants differ. For
instance, $C_{40} = M_{40} - 3 M_{20}^2$. 

Using the cumulants $C_{mn}$ we can easily construct the cumulants
$K_n$ of the total energy $H$, defined by the derivatives of $\ln Z$
with respect to $\beta$, see Eq.~\eqref{partfuncmodel}.  For example,
we have
\begin{eqnarray}
  K_2 &=& J^2 C_{20} + 2 J \kappa C_{11} + \kappa^2 C_{02}\,,\label{kncnm}\\
  K_3 &=& - \left( J^3 C_{30} + 3 J^2 \kappa C_{21} + 3 J \kappa^2 C_{12} +
  \kappa^3 C_{03}\right)\,,\nonumber
\end{eqnarray}
etc... Note that the specific heat is given by
$C_V=K_2/V$.

In the following, we consider periodic boundary conditions, which
preserve the duality property in finite-size systems. Using
Eq.~(\ref{dualityZ}), and taking the appropriate derivatives with
respect to $J$ and $\kappa$, we can obtain an infinite series of exact
relations among the expectation values $E_J,\,E_\kappa$ and the
cumulants $C_{mn}$ at $(J,\kappa)$ and at the corresponding
duality-transformed couplings $(J^\prime, \kappa^\prime)$,
cf. Eq.~(\ref{dualitymap}).  Along the self-dual line where
$(J,\kappa)=(J^\prime, \kappa^\prime)$, they turn into an infinite
series of exact relations among expectation values of cumulants
computed on the self-dual line.  The lowest-order cumulants satisfy
the relations
\begin{eqnarray}
&& E_{\kappa}+\sinh(2J)\,E_J-3\cosh(2J)L^3=0\,,
    \label{exadualrel1}\\
&& \sinh^2(2J)\,C_{20}-C_{02}-2\cosh(2J)\,E_{\kappa}+6L^3=0\,.\qquad
\label{exadualrel2}
\end{eqnarray}
Relations for higher-order cumulants are more cumbersome.  Neglecting
the regular terms arising from the second term of the r.h.s. of
Eq.~(\ref{dualitymap}), third-order cumulants satisfy the relations
  \begin{eqnarray}
&& C_{12} + \sinh(2J)\,C_{21} + 2 \cosh(2J) \,C_{11}  \approx
  0\,, \label{thirdcurel} \\
  && C_{03} + \sinh^3(2J) \,C_{30}+ 6   \cosh(2J)\,C_{02} +\nonumber\\
&&\quad + 2 [3 + \cosh(4J)]E_\kappa \approx 0 \,.
  \nonumber
\end{eqnarray}

  The scaling behavior of the cumulants $C_{nm}$ can be derived by
differentiating the asymptotic scaling relation
\begin{equation}
   F_{\rm sing}(J,\kappa,L) \approx L^{-d} f(x_1, x_2)\,, \qquad x_i =
   u_i L^{y_i}\,,
\label{scalF-finiteV}
\end{equation}
where we only keep the relevant RG contributions.  Note that the
duality relation (\ref{dualityZ}) for the free energy, and the duality
properties of $u_1$ and $u_2$, imply that
\begin{equation}
     f(-x_1,x_2) = f(x_1,x_2)\,.
\label{SF-funf}
\end{equation}
Introducing the derivatives
\begin{equation}
   f_{n,m}(x_1,x_2) = {\partial^{n+m} f(x_1,x_2) \over \partial x_1^n
     \partial x_2^m}\,,
\label{calcnmdef}
\end{equation}
the leading critical contribution is generally given by
\begin{eqnarray}
C_{nm}(J,\kappa,L) \approx u_{1,J}^n u_{1,\kappa}^m L^{(n+m)y_1}
f_{n+m,0}(x_1,x_2) \,,
\label{genscalingCnm}
\end{eqnarray}
where $u_{1,J}$ and $u_{1,\kappa}$ are the derivatives of $u_1$ with
respect to $J$ and $\kappa$. 
The cumulants of the total energy are expected to develop an
analogous scaling behavior, i.e.
\begin{eqnarray}
K_n(J,\kappa,L) \approx L^{n y_1} {\cal K}_{n}(x_1,x_2) \,.
\label{genscalingK}
\end{eqnarray}

Along the self-dual line $u_1=0$ the duality symmetry leads to some
cancellations, as a consequence of Eq.~(\ref{SF-funf}).  For $n+m$
even, the leading terms of the cumulants $C_{nm}$ are given by
\begin{eqnarray}
C_{nm}(J,\kappa,L) &\approx& u_{1,J}^n u_{1,\kappa}^m L^{(n+m)y_1}
f_{n+m,0}(0,x_2) \nonumber \\
&\approx& c^m L^{(n+m) y_1} f_{n+m,0}(0,x_2) \,.
\label{scaling-even}
\end{eqnarray}
Note that Eq.~(\ref{scaling-even}) is consistent with the exact
relations derived from duality, such as Eq.~(\ref{exadualrel2}).  For
$n+m$ odd, the relation (\ref{SF-funf}) implies that
\begin{equation}
f_{n+m,0}(0,x_2)= 0\,.
\label{cmnx10}
\end{equation}
Therefore, the leading scaling behavior is obtained by differentiating
once with respect to $u_2$. Thus, for odd $n+m$ we obtain
\begin{eqnarray}
  C_{nm} &\approx& L^{(n+m-1) y_1 + y_2} f_{n+m-1,1}(0,x_2) \times
\label{scaling-odd}\\
  &&\;\;\times (n\ u_{1,J}^{n-1} u_{1,\kappa}^{m} u_{2,J} +
  m\ u_{1,J}^n u_{1,\kappa}^{m-1} u_{2,\kappa}) \nonumber \\ &\approx& (m -
  n) c^m L^{(n+m-1) y_1 + y_2} f_{n+m-1,1}(0,x_2) \,,
  \nonumber
\end{eqnarray}
where $u_{2,J}$ and $u_{2,\kappa}$ are the derivatives of $u_2$ with
respect to $J$ and $\kappa$, respectively.  Using these asymptotic
behaviors along the self-dual line and the relations (\ref{kncnm}), we
can also derive the corresponding asymptotic FSS of the cumulants
$K_n$ of the total energy, which behave as
\begin{eqnarray}
  K_n & \approx & L^{ny_1}\, \widetilde{\cal K}_n(x_2) \quad {\rm for}
  \;\;{\rm even}\;\; n\,,
  \label{Hcumulants-scaling-even}\\
K_n &\approx& L^{(n-1)y_1+y_2} \, \widetilde{\cal K}_n(x_2)
  \quad {\rm for} \;\;{\rm odd}\;\; n\,.
\label{Hcumulants-scaling-odd}
\end{eqnarray}

It is also useful to consider combinations whose cumulants have
definite properties under duality transformations.  We
define
\begin{eqnarray}
A &=& H_J - \sinh(2\kappa)\,H_\kappa\,, \label{defA}\\
S &=& H_J + \sinh(2\kappa)\,H_\kappa\,.\label{defS}
\end{eqnarray}
Since 
\begin{eqnarray}
{\partial u_1\over \partial J} + \sinh(2\kappa)
{\partial u_1 \over \partial \kappa} 
&=& 0\,, \\
{\partial u_2\over \partial J} - \sinh(2\kappa)
{\partial u_2 \over \partial \kappa} &=& 0 \,,\nonumber
\end{eqnarray}
one can easily check that the cumulants $A_{n}$ of the operator $A$,
defined in Eq.~(\ref{defA}), do not receive contributions associated
with the scaling field $u_1$. Therefore, they generally scale as
\begin{eqnarray}
A_{n} \approx L^{n y_2} {\cal A}_n(x_1, x_2) \,,   
\quad  {\cal A}_n = 
(-2)^n  f_{0n}(x_1,x_2) \,.
\label{HnAsca}
\end{eqnarray}
The cumulants $S_{n}$ of the operator $S$ behave as
\begin{eqnarray}
  S_{n} \approx L^{n y_1} {\cal S}_n(x_1,x_2) \,,  \quad
{\cal S}_n = 
  2^n  f_{n0}(x_1,x_2) \,.
\label{HnSsca}
\end{eqnarray}
Along the self-dual line, however, this diverging behavior is not
observed for $n$ odd, since $f_{n0}(0,x_2)=0$, thus $S_n$ is expected
to diverge as $L^{(n-1) y_1}$.

We finally note that the above scaling equations assume that the
leading contribution is due to the singular part of the free
energy. However, contributions due to the regular free-energy term, of
order $L^d$, may provide the leading contribution for the lowest-order
cumulants, depending on the values of the RG exponents $y_1$ and
$y_2$.

\subsection{Multicritical field theory}
\label{sec3C}

The results of Sections \ref{sec3A} and \ref{sec3B} only rely on the
existence of a duality transformation and make no assumption on the
nature of the MCP. To go further and make more quantitative
predictions, it is crucial to understand the nature of the order
parameters. Along the finite-$J$ transition line that ends at $\kappa
= \infty$, the order parameter is expected to be a local function of
the $s_x$ fields, which should correspond to the Ising
magnetization. Of course, because of gauge invariance, any rigorous
definition requires the introduction of an appropriate gauge fixing,
which however would not change any gauge-invariant correlation
function (in Ref.~\onlinecite{BN-87} this approach has been used to obtain
rigorous results for the phase behavior of the U(1) Abelian-Higgs
model).  The order parameter for the ${\mathbb Z}_2$ gauge theory is
instead expected to be nonlocal and indeed the transition has a
topological nature. Apparently, this observation seems to indicate
that one cannot use standard symmetry arguments to understand the
critical behavior at the MCP, as they assume that the order parameters
are coarse-grained local functions of the microscopic fields.

We wish now to argue that, at the MCP (and only there), because of
duality, we can assume that both order parameters are local.  Strictly
speaking, duality is only a mapping of the Hamiltonian parameters, but
here we will enlarge its role and assume that duality provides a
mapping for all RG operators. Essentially, let us assume that we are
working in the infinite-dimensional space of Hamiltonians on which the
RG transformations act~\cite{Wegner}.  If we start from a ${\mathbb
  Z}_2$ gauge Hamiltonian, under RG transformations, we will generate
a flow towards a ${\mathbb Z}_2$ gauge-invariant fixed point, while
starting from the usual Ising model, we will observe a flow towards
the Wilson-Fisher ${\mathbb Z}_2$ fixed point. The existence of an
exact microscopic relation between the ${\mathbb Z}_2$ gauge model and
the Ising model allows us to conjecture that the two fixed points are
equivalent, with the same set of RG dimensions and operators. In other
words, there is a mapping (we call it duality) between all RG
operators at the different fixed points.  It is then plausible that
this duality transformation maps the local order parameter of the
Ising model to the nonlocal order parameter of the gauge model. The
mapping changes the Hamiltonian parameters, except on the self-dual
line, and therefore at the MCP.  Here, the mapping would imply the
equivalence of the local and of the nonlocal order parameters for the
same model.  Therefore, it seems reasonable to describe the
multicritical behavior in terms of two local quantities.  We thus
consider two different scalar fields $\varphi_1({\bm x})$ and
$\varphi_2({\bm x})$, associated with the two transition lines.

To derive a Lagrangian for the effective model, we note that the
theory should be invariant under a change of sign of both fields, so
that only even powers of each field are allowed.  Under these
conditions the LGW Hamiltonian is~\cite{LF-72,FN-74,NKF-74}
\begin{eqnarray}
{\cal H} &=& 
\frac{1}{2} \Bigl[ ( \partial_\mu \varphi_1)^2  + (
\partial_\mu \varphi_2)^2\Bigr] + 
\frac{1}{2} \Bigl( r_1 \varphi_1^2  + r_2 \varphi_2^2 \Bigr)
\nonumber\\
&&+
\frac{1}{4!} \Bigl[ v_1 \varphi_1^4 + v_2 \varphi_2^4 + 
                           2 w\, \varphi_1^2\varphi_2^2 \Bigr] \,.
\label{bicrHH} 
\end{eqnarray}
This model has been studied at length.  In the mean-field
approximation~\cite{LF-72,FN-74,NKF-74}, the field theory
(\ref{bicrHH}) admits a bicritical point analogous to the one
appearing in Fig.~\ref{phadiabis}. Moreover, if the transition is
continuous, it should belong to the $XY$ universality class
\cite{LF-72,FN-74,NKF-74,PV-02,CPV-03} thereby leading to an effective
enlargement of the symmetry from ${\mathbb Z}_2\oplus{\mathbb Z}_2$ to
O(2).

Several field-theoretical and numerical works have determined the
exponents $y_i$ entering the multicritical scaling ansatz
(\ref{freeen}), see, e.g., Refs.~\onlinecite{CPV-03,HV-11}. As shown in
Ref.~\onlinecite{CPV-03}, the leading exponents correspond to the RG
dimensions at the isotropic $XY$ fixed point of quadratic and quartic
perturbations that belong to different representations of O(2)
group. The leading RG exponent $y_1$ is associated with the quadratic
spin-two perturbation. The corresponding RG dimension
is~\cite{CPV-03,HV-11}
\begin{eqnarray}
y_1 = 1.7639(11)\,.
\label{y1XY}
\end{eqnarray}
The second largest exponent is associated with the spin-zero
quadratic operator, and is directly related to the 
correlation-length critical exponent at standard $XY$ transitions:
\begin{equation}
  y_2 = {1\over \nu_{xy}} = 1.4888(2)\,,
  \label{y2XY}
\end{equation}
where we used the estimate $\nu_{xy}=0.6717(1)$ (see, e.g.,
Refs.~\onlinecite{GZ-98,PV-02,CHPV-06,KP-17,Hasenbusch-19,CLLPSSV-20} for
theoretical results by various methods).  Using the above results, we
can estimate the crossover exponent,
\begin{equation}
\phi = y_1/y_2 = 1.1848(8)\,.
\label{phiXY}
\end{equation}
Scaling corrections at the multicritical $XY$ point are controlled by
the negative RG dimensions $y_i$. The most relevant ones are
\cite{CPV-03,CPV-00,HV-11,Hasenbusch-19}
\begin{eqnarray}
  y_3 &=& -0.108(6)\,,  \label{y3XY} \\
  y_4 &=& -0.624(10)\,,  \\
  y_5 &=& -0.789(4)\,,  \label{y5est}
\end{eqnarray}
which are related to the spin-4, spin-2, and spin-zero quartic
perturbations, respectively. Note that, at standard $XY$ transitions,
corrections decay as $L^{-\omega}$ with $\omega = - y_5 \approx 0.79$.  At
the MCP, scaling corrections decay much slower, as $L^{y_3}\approx
L^{-0.108}$, which may complicate the analysis of the universal
multicritical $XY$ behavior.  Moreover, corrections with any integer
combination of the subleading exponents are also expected, and thus
corrections $L^{ny_3}$ with $n=2,3,\ldots$ should also appear.

In the LGW approach the analogue of the duality mapping is the
exchange of the two fields ($\varphi_1\to\varphi_2$,
$\varphi_2\to\varphi_1$). The RG operators associated with the scaling
fields $u_i$ must have definite properties under these
transformations. The leading operator of RG dimension $y_1$ is odd
under the field exchange. This implies that $u_1$ is odd under the
simultaneous exchange of $r_1$, $r_2$ and of $v_1$, $v_2$.  In the
${\mathbb Z}_2$ gauge Higgs model this implies that the scaling field
$u_1$ is odd under duality, in agreement with the arguments presented
in Sec.~\ref{sec3A}. Analogously, we predict $u_2$ to be even, as
already discussed before. We can also predict the transformation
properties of the irrelevant scaling fields: $u_3$ and $u_5$ are even
functions under duality, while $u_4$ is odd. In particular, there are
no corrections with exponent $y_4$ on the self-dual line.

\section{Numerical results}
\label{numres}

In this section we report some numerical results supporting our
hypothesis of an emerging $XY$ multicriticality at the MCP, as
discussed in the previous sections.  For this purpose, we present FSS
analyses of MC simulations of the unitary-gauge model
(\ref{HiggsHug}), using a standard Metropolis upgrading of the
discrete spin link variables~\cite{Metropolis:1953am}.  We consider
cubic systems of size $L$ with periodic boundary conditions.

We perform simulations along the self-dual line $u_1=0$ and along the
line $u_2 = 0$, measuring the energy cumulants defined in
Sec.~\ref{sec3B}. We verify the predicted FSS behavior, using the RG
exponents $y_1 = 1.7639(11)$ and $y_2 = 1.4888(2)$ of the $XY$
universality class. We should remark that the observation of the
asymptotic scaling behaviors predicted by the multicritical $XY$
scenario is made difficult by the existence of several sources of
slowly decaying scaling corrections.  The leading ones decay very
slowly, as $L^{n y_3} \approx L^{- 0.108 n}$ with $n=1,2,\ldots$.
Then, we should consider terms decaying as $L^{-(y_1-y_2)} \approx
L^{-0.28}$ [they are absent on the self-dual line because of
  Eq.~(\ref{SF-funf})], as $L^{-2(y_1-y_2)} \approx L^{-0.55}$,
$L^{-y_4}\approx L^{-0.62}$ (they are absent along the self-dual
line), $L^{-y_5} \approx L^{-0.79}$.  For $m+n=2$ also the regular
background plays a role, giving rise to corrections of order
$L^{3-2y_1} \approx L^{-0.53}$.

\begin{figure}[tbp]
\includegraphics[width=0.95\columnwidth, clip]{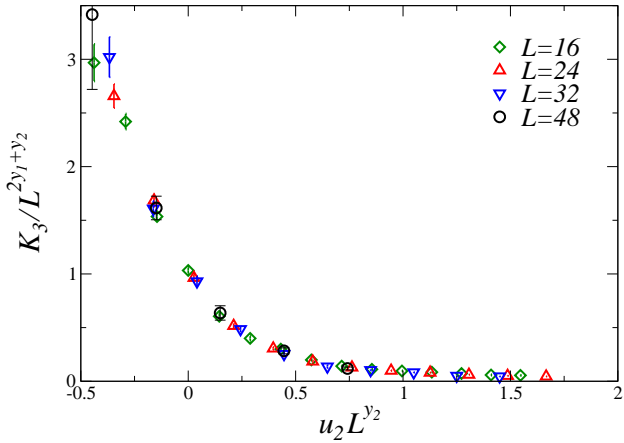}
\caption{Scaling behavior of the third cumulant $K_3$ of the
  Hamiltonian along the $u_1=0$ line as a function of $u_2 L^{y_2}$.
  We use the $XY$ exponents $y_1 = 1.7639$, $y_2=1.4888$ and
  $\kappa^{\star} = 0.7525$.}
\label{h3u10}
\end{figure}

\begin{figure}[t]
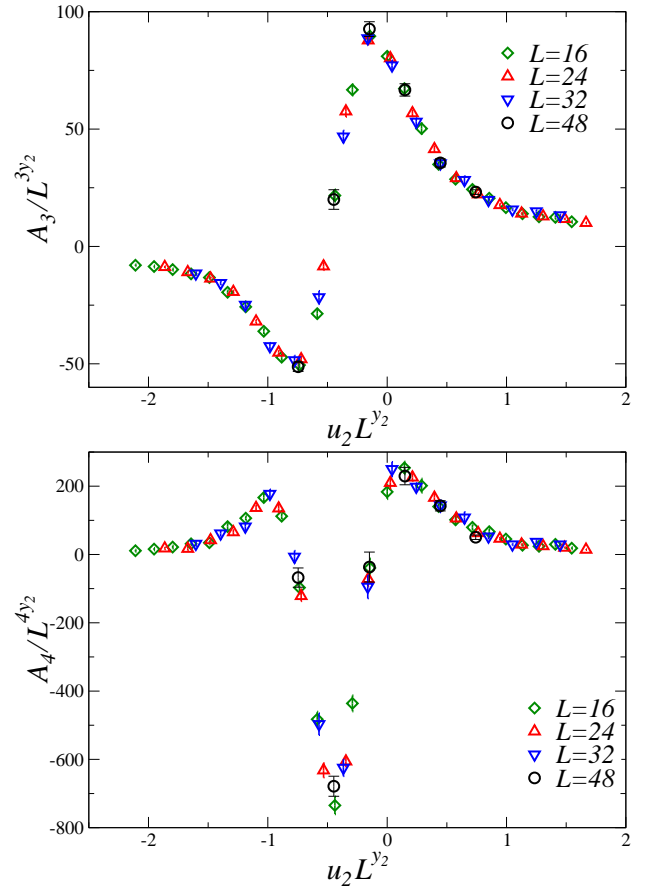

\includegraphics[width=0.95\columnwidth, clip]{u1zero_u2_H3A.eps}
\includegraphics[width=0.95\columnwidth, clip]{u1zero_u2_H4A.eps}
\caption{Scaling behavior of cumulants $A_{3}$ (top) and $A_{4}$ (bottom)
  along the $u_1=0$ line as a function of $u_2 L^{y_2}$.  We use the
  $XY$ exponents $y_1 = 1.7639$, $y_2=1.4888$ and $\kappa^{\star} =
  0.7525$.  }
\label{HAu10}
\end{figure}

Along the self-dual line the scaling field $u_1$ vanishes.  Thus,
according to the RG analysis reported in Sec.~\ref{sec3B}, we expect
that the asymptotic scaling behavior of the energy cumulants depends
on the FSS variable $x_2=u_2 L^{y_2}$.  Along the self-dual line the
numerical FSS analyses of the energy cumulants $K_n\,,A_n\,,S_n$ are
consistent with the predictions of the multicritical theory, see
Eqs.~(\ref{Hcumulants-scaling-even}) and
(\ref{Hcumulants-scaling-odd}) for the total energy, once the $XY$
values of the RG exponents reported in Eqs.~(\ref{y1XY}) and
(\ref{y2XY}) are used.  The most accurate estimate of the MCP point is
obtained by biased analyses of the third cumulant $A_{3}\sim L^{3y_2}$
of the operator $A$, see Eq.~(\ref{defA}), along the self-dual line,
using the $XY$ values for the exponents.  Fitting the data to
Eq.~(\ref{HnAsca}), we obtain
\begin{equation}
\kappa^\star = 0.7525(1)\,,\quad J^\star = 0.22578(5)\,.
\label{mcpco}
\end{equation}
This estimate of the MCP is consistent with the results reported in
Ref.~\onlinecite{SSN-21}.  The analysis of the other cumulants gives
consistent results.

The accuracy of the description in terms of the multicritical $XY$
predictions is demonstrated by the scaling plots of the data of the
cumulants using the $XY$ exponents and the estimates (\ref{mcpco}).
In Fig.~\ref{h3u10} we show data for the third cumulant $K_3$ of the
Hamiltonian, which is expected to scale with the power law $L^{2y_1 +
  y_2}$, cf. Eq.~(\ref{Hcumulants-scaling-odd}).  We observe a
reasonably good scaling: scaling corrections are hardly visible within
the statistical errors. Note that, according to the multicritical $XY$
scenario, one expects slowly decaying corrections with exponent
$|y_3|\approx 0.11$, cf. Eq.~(\ref{y3XY}). We do not observe them
here. In our range of values of $L$, $L^{y_3}$ varies only slightly,
and thus it is conceivable that they do not affect the divergent
behavior of the observables, but only the accuracy of the scaling
functions. In Fig.~\ref{HAu10} we report the scaling plots of $A_{3}$
and $A_{4}$.  Data are again in good agreement with the theoretical
predictions for their asymptotic scaling behavior,
cf. Eqs.~(\ref{HnAsca}).  We do not report the second cumulant
$A_{2}$, whose singular part should scale as $L^{2 y_2}$. Since $2 y_2
\approx 2.9775 < 3$, its behavior is dominated by the regular
contribution, that scales as the volume $L^3$.

\begin{figure}[t]
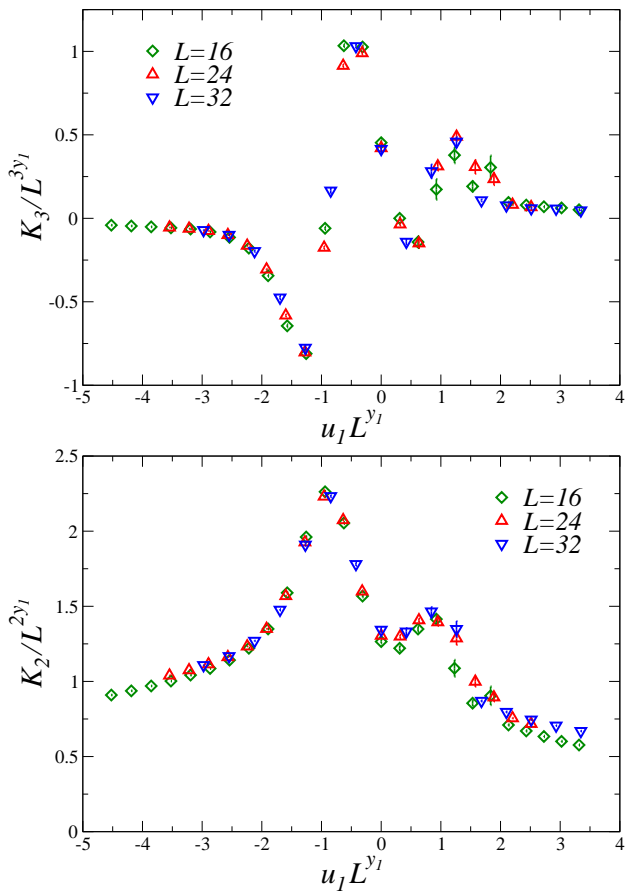

  \includegraphics[width=0.95\columnwidth, clip]{u2zero_u1_H3.eps}
  \includegraphics[width=0.95\columnwidth, clip]{u2zero_u1_Cv.eps}
  \caption{ Scaling plot of the second cumulant $K_2$ (bottom) 
    and of the third cumulant $K_3$ (top) of the Hamiltonian
    along the $u_2=0$ line, using the $XY$ exponent $y_1=1.7639$ and
    $\kappa^{\star}=0.7525$. Data confirm the 
    scaling prediction, Eq.~(\ref{knscau20}) .
    Notice that the error bars of $K_3$ for
    $u_1 L^{y_1}\gtrsim 1$ may be underestimated. In this 
    region of the phase diagram we observe an  increasing
    inefficiency of the MC algorithm.}
\label{h3u20}
\end{figure}

Beside checking the consistency of the numerical data 
with the multicritical $XY$ scenario, we can also perform unbiased 
fits, to determine $y_1$ and $y_2$.
If we fit the third and fourth cumulant of the Hamiltonian
(they should scale as $K_3\sim L^{2 y_1 + y_2}$ and $K_4\sim L^{4
  y_1}$, respectively) we obtain $2 y_1 + y_2 = 5.0(1)$ and $4 y_1 =
7.0(1)$, which are consistent with the predictions $2 y_1 + y_2
\approx 5.02$ and $4 y_1 \approx 7.06$.  The exponent $y_2$ can also
be estimated from $A_{n}$. We obtain $y_2 = 1.495(10)$ and $y_2 =
1.48(2)$ from $A_{3}$ and $A_{4}$, respectively.  The agreement with
the conjectured $XY$ values is quite good.

We also performed simulations along the $u_2=0$ line, see
Eq.~(\ref{u2def}), using the estimate $\kappa^{\star}=0.7525$ obtained
from the FSS analyses along the self-dual $u_1=0$ line. Along the
$u_2=0$ line, the asymptotic FSS of the cumulants of the total energy
is that given in Eq.~(\ref{genscalingK}), i.e.
\begin{equation}
K_n \approx L^{ny_1} {\cal K}_n(x_1,0)\,.
\label{knscau20}
\end{equation}
Note that, for $n$ odd, consistency with the FSS behavior along the
self-dual line, see Eq.~(\ref{Hcumulants-scaling-odd}), requires
${\cal K}_n(0,0) = 0$.  The data are plotted in Fig.~\ref{h3u20}, We
observe a nice collapse of the data, again fully supporting the
multicritical $XY$ scenario.

\begin{figure}[t]
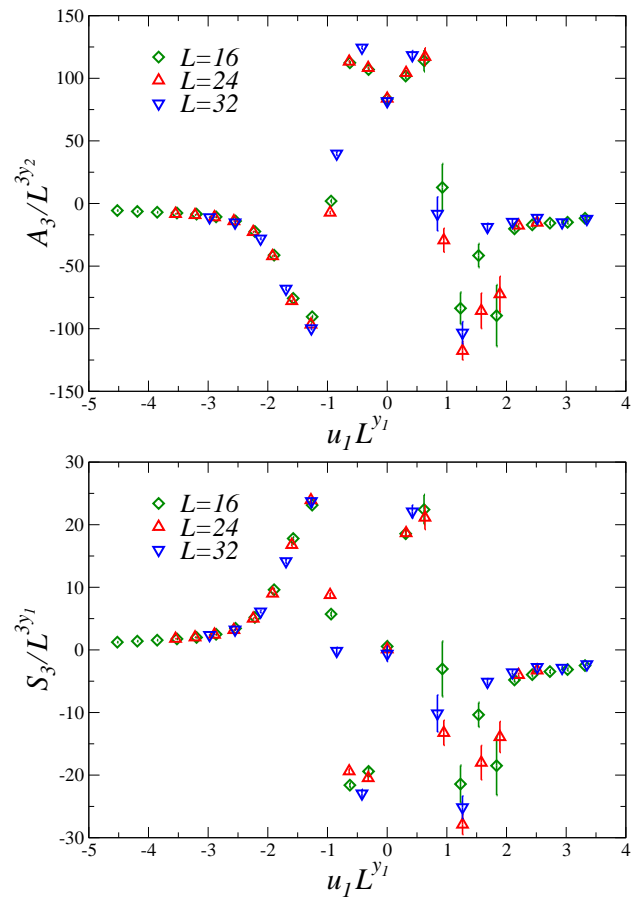

  \includegraphics[width=0.95\columnwidth, clip]{u2zero_u1_H3A.eps}
  \includegraphics[width=0.95\columnwidth, clip]{u2zero_u1_H3S.eps}
  \caption{ Scaling plot of the third cumulants $A_{3}$ (top) and
    $S_{3}$ (bottom) of the operators $A$ and $S$ along the $u_2=0$
    line, using the $XY$ exponents $y_1=1.7639$ and $y_2=1.4888$, and
    $\kappa^{\star}=0.7525$.  }
\label{AS3u20}
\end{figure}

Finally, we also check the scaling behavior of the third cumulants of
$A$ and $S$ along the $u_2=0$ line, in Fig.~\ref{AS3u20}. The scaling
behavior of the cumulants of $A$ is given in Eq.~(\ref{HnAsca}). It
depends on $f_{0n}(x_1,0)$ which is always an even function of
$x_1$. The data shown in the top Fig.~\ref{AS3u20} are definitely
consistent within the errors.  As for the cumulants of $S$, they scale
as reported in Eq.~(\ref{HnSsca}). Relation (\ref{SF-funf}) implies
that the odd (resp. even) cumulants are odd (resp. even) functions of
$x_1$.  Again, this is confirmed by the data shown in the bottom
Fig.~\ref{AS3u20}. In particular, the ratio $S_3/L^{3 y_1}$ is
consistent with zero at the critical point $x_1=0$.

Note that statistical errors of the MC simulations along $u_2=0$ line
increase significantly in the region $x_1 \gtrsim 1$.  The link update
algorithm for the model (\ref{HiggsHug}) becomes indeed less efficient
as $\kappa$ and $J$ are increased.  Autocorrelation times increase by
more than one order of magnitude, likely due to a different dynamic
regime related to the presence of relevant nonlocal configurations,
which are hardly modified by local moves.

The results we have presented here complement those reported in
Ref.~\onlinecite{SSN-21}, which were already providing a remarkable evidence
of the multicritical $XY$ behavior (although the authors were quite
skeptical on its interpretation in terms of a multicritical $XY$
behavior).  In particular, their estimates of the multicritical
exponents $y_1= 1.778(6)$ and $y_2=1.495(9)$ (other compatible, but
less precise, results were also reported in
Refs.~\onlinecite{VDS-09,DKOSV-11}) are in substantial agreement with the
$XY$ predictions (\ref{y1XY}) and (\ref{y2XY}). The small difference
in the estimate of $y_1$ can be easily explained by the very slowly
decaying scaling corrections predicted by the multicritical $XY$
scenario, that make a precise determination of the universal
asymptotic quantities very hard.  The leading one vanishes as
$L^{-0.108}$, so that, to reduce it by a factor of two, the lattice
size must be increased by a factor of 600, which is unattainable in
practice.

Overall, we believe that the numerical results presented in this
paper, and those already reported in Ref.~\onlinecite{SSN-21}, provide
strong evidence of the multicritical $XY$ scenario put forward in the
previous sections.

\section{Conclusions}
\label{conclu}

We study the multicritical behavior of 3D ${\mathbb Z}_2$ gauge Higgs
model at the MCP, where one first-order transition line and two
continuous Ising transition lines meet, as sketched in
Fig.~\ref{phadia}. The duality properties of the model play a key role
in the phase diagram, and in determining the main features of the
multicritical behavior at the MCP located on the self-dual line.

We exploit duality to identify the scaling fields associated with the
relevant RG perturbations at the MCP, and outline the corresponding
multicritical scaling behaviors.  Moreover, we present arguments
supporting the identification of the multicritical universality class
with the one controlled by the stable $XY$ fixed point of the 3D
multicritical LGW field theory (\ref{bicrHH}), with two competing
scalar fields associated with the continuous ${\mathbb Z}_2$
transition lines meeting at the MCP.  The $XY$ nature of the MCP
implies an effective enlargement of the symmetry of the multicritical
modes, to the continuous group O(2).

We have also reported numerical FSS analyses of several energy
cumulants.  The results are in good agreement with the theoretical
predictions based on the multicritical $XY$ scenario.  We believe that
our numerical results, together with those already reported in the
literature, see, in particular, Ref.~\onlinecite{SSN-21}, provide a strong
evidence in favor of the multicritical $XY$ scenario at the MCP.  Of
course, this picture calls for a deeper understanding of the
mechanisms that combine the local and nonlocal critical modes of the
${\mathbb Z}_2$ gauge Higgs model to give rise to the multicritical
$XY$ behavior, entailing an effective enlargement of the symmetry at
the MCP, to the continuous group O(2).

\end{document}